\documentclass[epj]{svjour}
\usepackage[dvips]{graphicx,rotating}

\begin{document}

\title{Reweighting of the form factors in exclusive $B \rightarrow X \ell 
\nu_{\ell}$ decays}


\author{D. C\^{o}t\'e, S. Brunet, P. Taras, B. Viaud} 

\institute{Laboratoire Ren\'e J. A. L\'evesque, Universit\'e de Montr\'eal, C. P.
6128, Succursale Centre-ville, Montr\'eal, QC H3C 3J7, Canada}

\date{May 27th, 2004}


\abstract{ A form factor reweighting technique has been elaborated to permit
relatively easy comparisons between different form factor models applied to 
exclusive $B \rightarrow X \ell \nu_{\ell}$ decays. The software tool developped 
for this purpose is described. It can be used with any event generator, three of 
which were used in this work: ISGW2, PHSP and FLATQ2, a new powerful generator. 
The software tool allows an easy and reliable implementation of any form factor 
model. The tool has been fully validated with the ISGW2 form factor hypothesis. 
The results of our present studies indicate that the combined use of the FLATQ2
generator and the form factor reweighting tool should play a very important role
in future exclusive $|V_{ub}|$ measurements, with largely reduced errors.}  

\PACS{{Form factors reweighting in exclusive semileptonic B decays}{}}

\maketitle


\section{Introduction}

Exclusive semileptonic $B \rightarrow X_{u} \ell \nu_{\ell}$ decays can be used 
to measure the magnitude of the CKM matrix element $V_{ub}$ as their branching 
fractions (B.F.) are related to $|V_{ub}|$ by the following relation:
\begin{equation}
\label{Vub}
|V_{ub}|=\sqrt{\frac{B.F.(B \rightarrow X_{u} \ell \nu_{\ell})}{x \cdot\tau_{B}}}
\end{equation}
where $\tau_{B}$ is the lifetime of the $B$ meson and $x$ is given by the 
relation : $\Gamma_{th} (B\rightarrow X_{u} \ell \nu_{\ell}) = x |V_{ub}|^2$, 
$\Gamma_{th}$ is the theoretical partial decay rate.

The study of exclusive semileptonic $B \rightarrow X_{u} \ell \nu_{\ell}$ decays 
offers some experimental advantages compared to an inclusive study of all 
$b \rightarrow u \ell \nu_{\ell}$ decays, such as the possibility of keeping a 
higher fraction of the phase space and permitting a better background rejection.
On the other hand, it deals with lower statistics and it is affected by large 
theoretical uncertainties arising from the calculation of the \emph{form factors}
describing the strong interaction effects on the hadronization of the 
different $X_{u}$ final states. These uncertainties in the form factors lead to 
different predictions of the shape of the differential decay rate which, in turn,
yield different predictions for the momentum spectrum of the lepton $\ell$, the
meson $X_u$ and its daughters. The subsequent varying efficiencies for the 
experimental cuts lead to uncertainties in the measured branching fractions. The 
theoretical uncertainties in the form factors also affect the values of $x$. The 
resulting uncertainty in $x$ is the largest source of uncertainty in the 
determination of $|V_{ub}|$ from exclusive branching fraction measurements. 

In addition, in many analyses (e.g. those performed in BaBar), the simulated 
inclusive lepton spectrum of B decays does not agree with data. Since the $B 
\rightarrow D \ell \nu_{\ell}$ and $B \rightarrow D^* \ell \nu_{\ell}$ are the 
most abundant of all $B$ decays, it is likely that at least part of the 
disagreement could arise from a wrong theoretical input to the simulation for 
these decays. Again, the most likely source of error comes from the form factors 
for these decays. Since the theoretical form factor predictions cover a rather 
large range, it is necessary to establish experimentally which theory best 
describes the data. This requires the possibility of varying the theoretical 
assumptions at the simulation level.

A tool to be described in this paper has been created for this purpose. It will 
permit to switch easily between various form factor hypotheses and/or to vary the
parameters of a given hypothesis, within a full standard Geant4 Monte Carlo (MC) 
simulation framework. The basic principle of this new tool is to generate events 
and run the full simulation and reconstruction sequence only once, with a given 
form factor hypothesis. Subsequently, the events thus generated are reweighted at
the ntuple level with the values provided by a different form factor hypothesis.

The relevant formulas and basic principles of the form factor reweighting 
technique are presented in Sect. \ref{sec:ScalarAndVectors}. In Sect. 
\ref{sec:software}, the structure of the software tool developped for this 
purpose and how to use it are described. In that section, we also show how to 
incorporate new form factor hypotheses in the tool. Several histograms 
\footnote{A far more extensive document is available \protect\cite{cote} from the
authors. This document presents a large number of useful relations, part of the 
$C^{++}$ sofware tool developped in this work as well as many more histograms.} 
that demonstrate that the tool is working properly are shown in Sect. 
\ref{sec:validation}. A sample of kinematical distributions calculated with 
different form factor models are compared in Sect. \ref{multimodels}. The 
conclusions are given in Sect. \ref{con}.

\section{Technique for form factors reweighting}
\label{sec:ScalarAndVectors}

\subsection{Pseudo-scalar versus vector mesons}

The exclusive $B \rightarrow X_{u} \ell^{+} \nu_{\ell}$ decays \footnote{Charge
conjugation is implied throughout this paper, unless explicitly stated otherwise.
} which could be studied with this technique are: $B^{+} \rightarrow \pi^{0}/
\eta/\eta'/\rho^{0}/\omega \ell^{+} \nu_{\ell}$ and $B^{0}\rightarrow \pi^{-}/
\rho^{-}\ell^{+} \nu_{\ell}$. As will be shown in Sect. \ref{DiffDec}, the 
differential decay rates are different for pseudo-scalar and vector 
mesons. The $B^{0}$, $B^{\pm}$, $\pi^{\pm}$, $\pi^{0}$, $\eta$ and $\eta'$ mesons
are all pseudo-scalar particles, while the $\rho^{\pm}$, $\rho^{0}$ and $\omega$ 
mesons are vector particles. As well, the most abundant $B \rightarrow X_c
\ell^{+} \nu_{\ell}$ decays, $B \rightarrow D \ell^{+} \nu_{\ell}$ and 
$B\rightarrow D^* \ell^{+} \nu_{\ell}$, involve $D$ pseudo-scalar mesons and 
$D^{*}$ vector mesons.

\subsection{Kinematics of semileptonic decays}
\label{sec:Kin}

\begin{figure}[ht]
\begin{center}
\mbox{
\resizebox{0.5\textwidth}{!}{\includegraphics{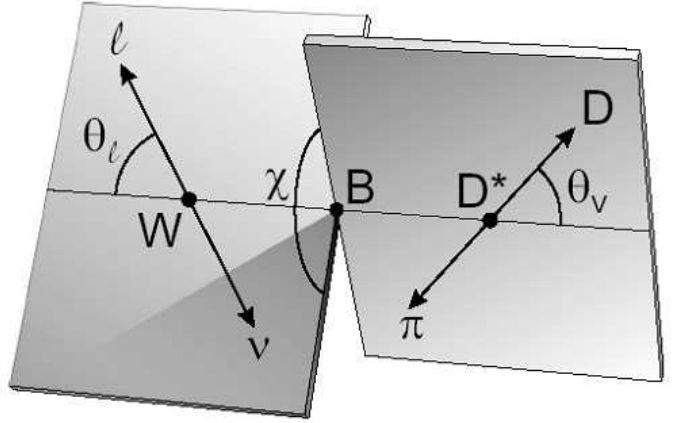}}}
\caption[Illustration of $\theta_{\ell}$, $\theta_{V}$ and $\chi$]
{ \label{fig1} Definition of the angles $\theta_{\ell}$, $\theta_{V}$ and
$\chi$ in the case of the $B$ to vector semileptonic decay $B \rightarrow
D^* \ell \nu$, $D^* \rightarrow D \pi$. $\theta_\ell$ is the helicity angle of 
the $W^{\pm}$ boson, the angle between the direction of the $W$ boson boosted in
the $B$ frame and the direction of the lepton boosted in the $W^{\pm}$ frame 
(where the $\ell$ and $\nu$ are emitted back to back). $\theta_V$ is the helicity
angle of the X meson, here the angle between the direction of the $D^*$ meson 
boosted in the $B$ frame and the direction of the $D$ meson boosted in the $D^*$ 
frame (where the $D$ and $\pi$ mesons are emitted back to back). $\chi$ is the 
angle between the $\vec{W} \times \vec{\ell}$ plane and the $\vec{D^*} \times 
\vec{D}$ plane.}
\end{center}
\end{figure}

A semileptonic $B \rightarrow X \ell \nu$ decay is generally described by the 
following process. The $B$ meson first decays into a virtual $W^{\pm}$ boson and 
an $X$ meson which are emitted back to back in the $B$ frame. The virtual 
$W^{\pm}$ boson then decays to a lepton and a neutrino which are emitted back to 
back in the $W^{\pm}$ frame, while the $X$ meson decays in various ways. The 
kinematics of such semileptonic decays can be completely described by three 
angles: $\theta_{\ell}$, $\theta_{V}$ and $\chi$ defined in Fig. \ref{fig1} and 
by $q^2$, the invariant mass squared of the virtual $W^{\pm}$ boson. In terms of
4-momenta:
\begin{equation}
q^2 = (p_\ell + p_\nu)^2 = (p_B - p_X)^2
\end{equation}
The four variables are totally uncorrelated. In the $B$ frame, $q^2$ is also 
given by:
\begin{equation}
\label{q2}
q^2 = m_B^2 + m_X^2 - 2m_B E_X 
\end{equation}

\noindent
where $E_X$ is the total energy of the $X$ meson. 
It is also of interest to note that the magnitude of the 3-momentum $\vec{p_X}$ 
and $q^2$ are uniquely related by Eq. \ref{pXq2} in the $B$ frame:
\begin{equation}
\label{pXq2}
|\vec{p_X}| = \sqrt{\frac{(m_B^2 + m_X^2 - q^2)^2}{4m_B^2} - m_X^2}
\end{equation}

\subsection{Form factors}
\label{sec:FF}

The matrix element of a semileptonic $B \rightarrow X \ell \nu_{\ell}$ decay 
can be written as \protect\cite{BurchatRichman}:
\begin{equation}
\label{matrix}
M(B \rightarrow X \ell \nu_{\ell}) = -i \frac{G_F}{\sqrt{2}}V_{xb}L^{\mu}H_{\mu}
\end{equation}
where $G_F$ is the weak interaction's Fermi constant, $V_{xb}$ is either $V_{ub}$
or $V_{cb}$ depending on the final state meson, $L^{\mu}$ is the leptonic current
and $H_{\mu}$ is the hadronic current. The leptonic current is well-known and can
be calculated precisely using perturbation theory.

The hadronic current accounts for the strong interactions between quarks and 
gluons, and thus for the hadronization of the final state quarks into a {\it X} 
meson. The hadronization of a $\bar{u}$ quark and a $d$ \emph{spectator} quark 
into a $\pi^{-}$ meson in a $B^0 \rightarrow \pi^{-} \ell^{+} \nu_{\ell}$ decay 
is a good example of such a situation. In all these processes which involve the 
exchange of soft gluons, the strong interaction coupling constant, 
$\alpha_s(\mu)$, is too large to allow the use of perturbative calculation 
techniques. Thus, even though the {\it structure} of hadronic currents is well 
known, such currents cannot be computed directly. However, they can be
parametrized in terms of a small number of so-called universal Isgur-Wise 
functions, or \emph{form factors} \protect\cite{isgw2}. To compute the form 
factors, it is necessary to use either non-perturbative calculation techniques 
such as lattice QCD \protect\cite{FNAL01}, or approximations of QCD. The 
approximation techniques
can themselves be split into two categories: 1) those, such as the Light Cone Sum
Rules (LCSR) \protect\cite{ball98}\protect\cite{ball01}, which are identical to 
QCD at some extreme limits but are a good approximation of QCD in a restricted 
but known kinematic range and 2) those, such as ISGW2 \protect\cite{isgw2} which,
instead of QCD, use approximate wave functions based on quark models for the 
mesons. The values of the form factors extracted from experimental data can then 
be confronted with various models or used on their own.

In terms of form factors, the hadronic current is given by a different expression
depending on whether the $B$ meson decays to a pseudo-scalar or to a vector 
meson final state. In both cases however, the expression for the hadronic current
can be simplified in the limit \protect\cite{BurchatRichman} of a massless 
lepton, assumed to be true when $\ell = e$ or $\mu$ (see Sect. \ref{massless} for
an estimate of the effect of this approximation). 

For decays such as $\bar{B}^0 \rightarrow \pi^{+} \ell^{-} \bar{\nu}_{\ell}$ 
where $\pi^{+}$ is a pseudo-scalar meson, the hadronic current is written,
in the limit of a massless lepton as \protect\cite{BurchatRichman}:
\begin{equation}
\label{hads}
H^{\mu} = \langle \pi^{+}(p')|u \gamma^\mu b|\bar{B^0}(p) \rangle = 
f^{+}(q^2)(p + p')^{\mu} 
\end{equation}
where $q = p - p'$ and $f^{+}(q^2)$ is the form factor describing the
non-perturbative QCD effect.

For decays such as $\bar{B}^0 \rightarrow \rho^{+} \ell^{-} \bar{\nu}_{\ell}$ 
where $\rho^{+}$ is a vector meson, the hadronic current is written, in 
the limit of a massles lepton as \protect\cite{BurchatRichman}:
\begin{eqnarray}
\label{hadv}
H^{\mu} &=& \langle \rho^{+}(p',\epsilon)|u \gamma^{\mu}(1-\gamma_5) b
|\bar{B^0}(p) 
\rangle \nonumber \\
\\
&=& \frac{2i\epsilon^{\mu \nu \alpha \beta}}{m_B + m_{\rho}}
\epsilon^{*}_{\nu}p'_{\alpha}p_{\beta}V(q^2) 
-(m_B + m_{\rho})\epsilon^{*\mu}A_1(q^2) \nonumber \\
&+& \frac{\epsilon^{*} \cdot q}{m_B + m_{\rho}}(p+p')^{\mu}A_2(q^2) \nonumber
\end{eqnarray}
where $A_1(q^2)$, $A_2(q^2)$ and $V(q^2)$ are the three form factors describing 
the non-perturbative QCD effect. 

The equations for these two hadronic currents as well as their expressions
without the massless lepton approximation are discussed in various papers, for 
example in Ref. \protect\cite{BurchatRichman}. 

It is important to note that the leptonic current and the {\it structure} of the 
hadronic currents (Eqs. \ref{hads} and \ref{hadv}) follow directly
from Lorentz invariance and are thus model-independent. The theoretical
uncertainties in exclusive semi-leptonic decay analyses are only due to the 
uncertainties in the knowledge of the form factor(s): $f^{+}(q^2)$, $A_1(q^2)$, 
$A_2(q^2)$ and $V(q^2)$ which are model-dependent. These factors nominally depend
only on a single variable: $q^2$. However, in the case of vector meson decays, 
interference effects between the $A_1(q^2)$, $A_2(q^2)$ and $V(q^2)$ form factors
introduce an additional model dependence for the angular differential decay rates
\protect\cite{cote}.

\subsection{Differential decay rates}
\label{DiffDec}

The differential and total decay rates are the observable manifestations of the 
underlying leptonic and hadronic currents. According to quantum field theory, 
the decay rate is given by a squared matrix element containing a combination of
leptonic and hadronic currents. Just like for the structure of the current 
equations (Sect. \ref{sec:FF}), the {\it structure} of the differential and total
decay rate equations is also considered to be model-independent. Only the form
factors appearing in the rates give rise to theoretical uncertainties in 
exclusive semi-leptonic decay analyses \protect\cite{cote}. These will be 
investigated.

Our investigation will be greatly simplified by the use of a technique to 
reweight the form factors among the various hypotheses under study. As will be 
shown in Sect. \ref{Rew}, this reweighting is equivalent to a reweighting of the 
differential decay rates i.e. of the probabilities of generating an event. The 
total decay rate $\Gamma$ does have a large effect \protect\cite{cote} on the 
model dependency of $B \rightarrow X \ell \nu$ analyses but is not used in the 
context of form factor reweighting.  

In the limit of a massless lepton, the differential decay rate of semileptonic B 
decays to a pseudo-scalar meson is given by \protect\cite{BurchatRichman}:
\begin{equation}
\label{dGs}
\frac{d\Gamma(B \rightarrow S \ell^{+} \nu_{\ell})}{dq^{2}d\cos\theta_{\ell}
d\cos\theta_{V}d\chi} = |V_{xb}|^{2} \frac{G_F^2}{128\pi^{4}}|\vec{p_{S}}|^3
\sin^2\theta_{\ell}\sin\theta_V|f^{+}(q^2)|^2
\end{equation} 
where $q^{2}$, $\theta_{V}$, $\theta_{\ell}$ and $\chi$ have been defined in 
Sect. \ref{sec:Kin}, $\vec{p_{S}}$ is the 3-momentum of the final state 
pseudo-scalar meson in the $B$ frame, $f^{+}(q^2)$ is the QCD form factor 
described in Sect. \ref{sec:FF} and $V_{xb}$ is either $V_{ub}$ or $V_{cb}$ 
depending on the final state meson. It is often practical to use an expression 
where two or three of the angles are integrated out in which case Eq. \ref{dGs} 
becomes \protect\cite{Gibbons}:
\begin{equation}
\label{dGss}
\frac{d\Gamma(B \rightarrow S \ell^{+} \nu_{\ell})}{dq^{2}d\cos\theta_{\ell}} 
= |V_{xb}|^{2} \frac{G_F^2}{32\pi^{3}}|\vec{p_{S}}|^3\sin^2\theta_{\ell}
|f^{+}(q^2)|^2
\end{equation} 
\begin{equation}
\label{dGsss}
\frac{d\Gamma(B \rightarrow S \ell^{+} \nu_{\ell})}{dq^2} 
= |V_{xb}|^{2} \frac{G_F^2}{24\pi^{3}}|\vec{p_{S}}|^3|f^{+}(q^2)|^2
\end{equation} 
 
Also, in the limit of a masless lepton, the differential decay rate of 
semileptonic B decays to a vector meson is \protect\cite{BurchatRichman}:
\begin{eqnarray*}
&&\frac{d\Gamma(B \rightarrow V \ell^{+} \nu_{\ell})}
{dq^{2}d\cos\theta_{\ell}d\cos\theta_{V}d\chi} 
 =  |V_{xb}|^{2} \frac{3G_F^2|\vec{p_{V}}|q^2}{8(4\pi)^{4}m_B^2} \\
&\times& \left\{  \begin{array}{l} (1-\cos\theta_{\ell})^{2}\sin^2\theta_V|H_{+}
(q^2)|^2 \\
+ (1+\cos\theta_{\ell})^{2}\sin^2\theta_V|H_{-}(q^2)|^2 \\
+ 4\sin^2\theta_{\ell}\cos^2\theta_V|H_{0}(q^2)|^2 \\
- 4\sin\theta_{\ell}(1-\cos\theta_{\ell})\sin\theta_V\cos\theta_V\cos\chi 
H_{+}(q^2)H_{0}(q^2) \\
+ 4\sin\theta_{\ell}(1+\cos\theta_{\ell})\sin\theta_V\cos\theta_V\cos\chi 
H_{-}(q^2)H_{0}(q^2) \\
- 2\sin^2\theta_{\ell}\sin^2\theta_V\cos2\chi H_{+}(q^2)H_{-}(q^2) \\
\end{array} \right\}
\end{eqnarray*}\begin{equation}\label{dGv}\end{equation}
where $\vec{p}_{V}$ is now the 3-momentum of the final state vector meson in
the $B$ frame. The functions $H_{+}(q^2)$, $H_{-}(q^2)$ and $H_{0}(q^2)$ are 
known as the helicity amplitudes of the vector meson. They are related 
\protect\cite{cote} to the QCD form factors $A_1(q^2)$, $A_2(q^2)$ and $V(q^2)$ 
described in Sect. \ref{sec:FF}. Integrating out the angles, Eq. \ref{dGv} 
becomes:
\begin{eqnarray}
&&\frac{d\Gamma(B \rightarrow V \ell^{+} \nu_{\ell})}{dq^2} = 
|V_{xb}|^{2} \frac{G_F^2|\vec{p_{V}}|q^2}{96\pi^{3}m_B^2} \times \nonumber \\ 
&& \;\;\;\;\;\;\;\;\;\;\;\;\;\;\;\; 
\left(|H_{+}(q^2)|^2+|H_{-}(q^2)|^2+|H_{0}(q^2)|^2\right) 
\end{eqnarray}

As can be seen from Eqs. \ref{dGs} and \ref{dGv}, the form factors introduce a 
model dependence in the prediction of the shape of the differential decay rates 
of both pseudo-scalar and vector meson decays. This will be shown explicitly in 
Sect. \ref{multimodels}. With standard procedures, this theoretical uncertainty 
in the shape of the differential decay rate leads to a ``theoretical'' 
uncertainty in the efficiency of the experimental cuts, and thus on the measured 
branching fraction. 
 
\subsection{Reweighting the probabilities of generating events among various form
factor models}
\label{Rew}

 A useful feature of the form factor reweighting tool is that events are 
generated, fully simulated and reconstructed only once, using a given form factor
hypothesis. The probabilities of generating events are then reweighted to any 
other form factor model at the ntuple level. Such a technique presents enormous 
advantages in terms of flexibility, time, CPU resources and disk space required,
compared to generating, fully simulating and reconstructing separate data samples
for each form factor model to be investigated.

Given the events generated with a certain probability by a specific form factor
model G, the probabilities of generating events according to a different form
factor hypothesis O, are obtained by applying a weight $w$ to the probabilities
of generating the events of type G. For a pseudo-scalar meson decay, the weights 
are defined as: 
\begin{equation} 
w = \frac{(\frac{d\Gamma(B \rightarrow S \ell \nu)}{dq^{2}d\cos\theta_{\ell}})
_O}{(\frac{d\Gamma(B \rightarrow S \ell \nu)}{dq^{2}d\cos\theta_{\ell}})_G} 
\end{equation}
and for a vector meson decay as:
\begin{equation} 
w = \frac{(\frac{d\Gamma(B \rightarrow V \ell \nu)}{dq^{2}d\cos\theta_{\ell}
d\cos\theta_{V}d\chi})_O}{(\frac{d\Gamma(B \rightarrow V \ell \nu)}{dq^{2}
d\cos\theta_{\ell}d\cos\theta_{V}d\chi})_G} 
\end{equation}

In this work, three different generators were used to calculate the initial 
probabilities for generating events. These probabilities were then reweighted to 
other form factor models. The generators are: 
\begin{itemize}
\item
ISGW2: This generator, based on a quark model calculation \protect\cite{isgw2}, 
is extensively used in BaBar, Belle and CLEO. It is used for the simulation of 
generic BBbar events including that of the $B\rightarrow X_u \ell \nu$ decays.
The differential decay rates are computed in this hypothesis using Eqs. 10 and 12
and the form factors given in Ref. \protect\cite{isgw2}\footnote{It should be 
mentionned that there are typographical errors in  Ref. \protect\cite{isgw2}. 
These have been corrected in our work.}.
\item
PHSP: This \emph{PHase SPace} generator has been used for several decays in our 
work. It generates events with equal probability in all points of the phase 
space. In the context of this generator, the differential decay rate is given by
the relation: $\frac{d\Gamma(B \rightarrow X \ell \nu)}
{dq^{2}d\cos\theta_{\ell}d\cos\theta_{V}d\chi} = constant \times|\vec{p}_X|$.
This means that the generated $\cos\theta_l$, $\cos\theta_V$ and $\chi$ 
distributions are flat while the differential decay rate decreases almost 
linearly with $q^2$ (see Eq. \ref{pXq2} and Fig. \ref{fig4}).
\item
FLATQ2: This new generator \protect\cite{cote} has recently been implemented in 
the BaBar software. It defines the probability of each event as the probability 
given by the PHSP generator divided by the value of $|\vec{p}_{X}|$ in the B 
frame, with a cut-off at $|\vec{p}_{X}| > 0.01$ $GeV/c$. In this case, the 
differential decay rate is given by the relation: $\frac{d\Gamma(B \rightarrow X 
\ell \nu)}{dq^{2}d\cos\theta_{\ell}d\cos\theta_{V}d\chi} = constant$. This means 
that the generated $\cos\theta_l$, $\cos\theta_V$, $\chi$ and $q^2$ distributions
are all flat i.e. all the events are generated with an equal probablility 
throughout this 4-dimensional space. Of the three generators, this one is the 
most useful to extract the form factors of $B \rightarrow X \ell \nu$ decays. In 
particular, the fact that events are generated with an equal probability for the 
complete $q^2$ range is useful to evaluate the efficiency of the experimental 
cuts, especially at high $q^2$ where most models predict very few events. The 
high $q^2$ events are of the utmost importance in the study of lattice QCD 
results. 
\end{itemize}

In cases where the initial distributions are generated with e.g. the ISWG2 
generator, the distributions for any other form factor model, e.g. the LCSR one, 
will be obtained by applying, in the case of a $B \rightarrow S \ell \nu$ decay, 
the following weights to the ISGW2 distributions:
\begin{eqnarray}
 w &=& \frac{|V_{xb}|^{2} \frac{G_F^2}{32\pi^{2}}|\vec{p}_S|^3\sin^2
\theta_{\ell}|f^{+}_{LCSR}(q^2)|^2}
{|V_{xb}|^{2} \frac{G_F^2}{32\pi^{2}}|\vec{p}_{S}|^3\sin^2\theta_{\ell}
|f^{+}_{ISGW2}(q^2)|^2} \\
&=& \frac{|f^{+}_{LCSR}(q^2)|^2}{|f^{+}_{ISGW2}(q^2)|^2}
\end{eqnarray}

If the same $B \rightarrow S \ell \nu$ decay distributions are generated 
initially with the FLATQ2 generator, the event-by-event weight has, in this case,
a rather simple form:
\begin{equation}
 w = \sin^2\theta_{\ell}|\vec{p}_{S}|^3|f^{+}_{LCSR}(q^2)|^2 
\end{equation}

\section{The form factor reweighting software tool}
\label{sec:software}

\subsection{Outline}

The form factor reweighting software has been written in $C^{++}$. The code is 
\emph{practically} self-contained, so that it can easily be used outside a 
specific framework (in that case, the CLHEP libraries need to be included). The 
code is written with an object oriented structure so that its different sections 
are independent of each other, each section being a separate class. For example, 
the section for computing the kinematic variables of Sect. \ref{sec:Kin} is 
independent of the one for computing the differential decay rate formulas and 
weights  which in turn is independent of the one for computing the form factors 
of the different models. This structure of separate classes having as input 
arguments either the LorentzVectors in the LAB frame or the output objects of the
other classes in the reweighting software, controlled by simple user interfaces, 
yields a high degree of versatility. This design should allow an easy and 
reliable implementation of expected new form factor models. 

\subsection{Software architecture}
\label{arch}

The form factor reweighting software consists of two separate tools: XSLKin and 
XSLEvtFFWeight, built on three inheritance levels, each made up of one or more
classes. The top level class of each tool is the user's interface from which all 
the lower classes are inheriting, directly or indirectly. All variables or
functions needed by the user are declared  (known as ``pure virtual functions'') 
at this level and defined and used for computation at the lower levels. The 
software architecture is shown in Figs. \ref{fig2} and \ref{fig3}. The three 
level structure of the XSLKin class diagram (Fig. \ref{fig2}) will probably be 
simplified in a future version of the sofware. On the other hand, the three level
structure of the XSLEvtFFWeight tool (Fig. \ref{fig3}) is very efficient in 
computing the reweighting among the various models to be investigated, as will be
explained in Sect. \ref{newff}.

\subsubsection{The XSLKin tool}
\label{XSLKin}
\begin{figure}[ht]
\begin{center}
\mbox{
\resizebox{0.15\textwidth}{!}{\includegraphics{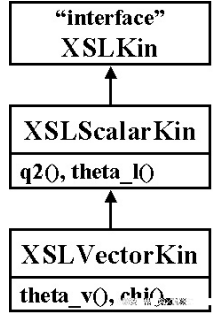}}}
\caption[XSLKin class diagram]{ \label{fig2} XSLKin class diagram. The arrows
mean an inheritance.}
\end{center}
\end{figure}

The XSLKin tool computes the kinematic variables described in Sect. 
\ref{sec:Kin}, 
namely: $q^{2}$, $\theta_{\ell}$, $\theta_{V}$, $\chi$. The objects of this 
tool are used as input arguments by the XSLEvtFFWeight tool to compute each
event weight. The XSLKin tool can also be used as a standalone for other 
physics analyses. From the $C^{++}$ point of view, the tool's structure is 
quite simple, as shown in Fig. \ref{fig2}. 

The XSLKin user's interface is used to declare all the kinematic variables needed
by users of this tool. The variables are defined and computed at the second 
and third levels. The second level class, named XSLScalarKin, is used to compute 
the kinematic variables ($q^{2}$ and $\theta_{\ell}$) characterizing a B to 
pseudo-scalar meson decay. The third level class, named XSLVectorKin, is used to 
compute the two additional kinematic variables ($\theta_{V}$ and $\chi$) needed 
to describe fully a B to vector meson decay. The XSLVectorKin class already 
contains the values of $q^{2}$ and $\theta_{\ell}$ inherited from the 
XSLScalarKin class.

This tool was created with an interface structure to allow the eventual 
addition of a second set of classes to compute the kinematic variables 
differently. It is not clear at present if this second set of classes will ever 
be used.

\subsubsection{The XSLEvtFFWeight tool}
\label{XSLEvtFFWeight}

\begin{figure}[ht]
\begin{center}
\mbox{
\resizebox{0.5\textwidth}{!}{\includegraphics{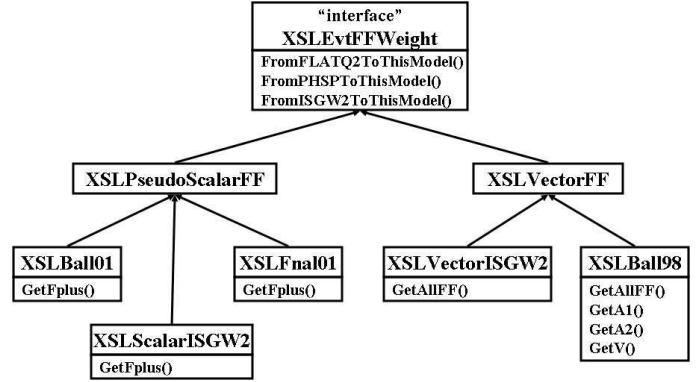}}}
\caption[XSLEvtFFWeight class diagram]{ \label{fig3} XSLEvtFFWeight 
class diagram. The arrows mean an inheritance.}
\end{center}
\end{figure}

The XSLEvtFFWeight tool architecture is shown in Fig. \ref{fig3}. 
The top level user's interface of this tool is used to declare the 
functions needed by the tool's user for reweighting an event. The second level of
the diagram contains the classes that inherit from the XSLEvtFFWeight interface. 
In these classes, the functions declared in the interface are written out 
explicitly for the cases to be investigated i.e. for the B to pseudo-scalar meson
decays in the XSLPseudoScalarFF class and for the B to vector meson decays in 
the XSLVectorFF class. These functions are used only to calculate the 
weight required to reweight the probability of generating an event from a given 
generator to a given form factor model. The form factors themselves as well as 
the kinematic parameters are computed in other classes. At the second level, the 
form factors are declared as pure virtual functions (GetFplus() and GetAllFF()) 
to be defined and computed at the third level. The third level classes are then 
used to compute these GetFplus() or GetAllFF() virtual functions of the second 
level i.e. the $f^{+}(q^2)$ or $A_1(q^2)$, $A_2(q^2)$ and $V(q^2)$ form factor 
values as a function of $q^2$ for various models. 

\subsection{How to implement new form factor models}
\label{newff}

The XSLEvtFFWeight tool has been created with a three-level structure to make it
efficient in computing the reweighting among various form factor models: the
form factors are modified at the third level while the other two levels
remain the same independently of the form factors used. The third level class 
inherits from either the XSLPseudoScalarFF or XSLVectorFF class. The 
mathematical functions required to compute the form factor(s) as a function of 
$q^2$ are inserted in this class. 

\begin{figure*}[t]
\begin{center}
\mbox{
\resizebox{0.5\textwidth}{!}{\includegraphics{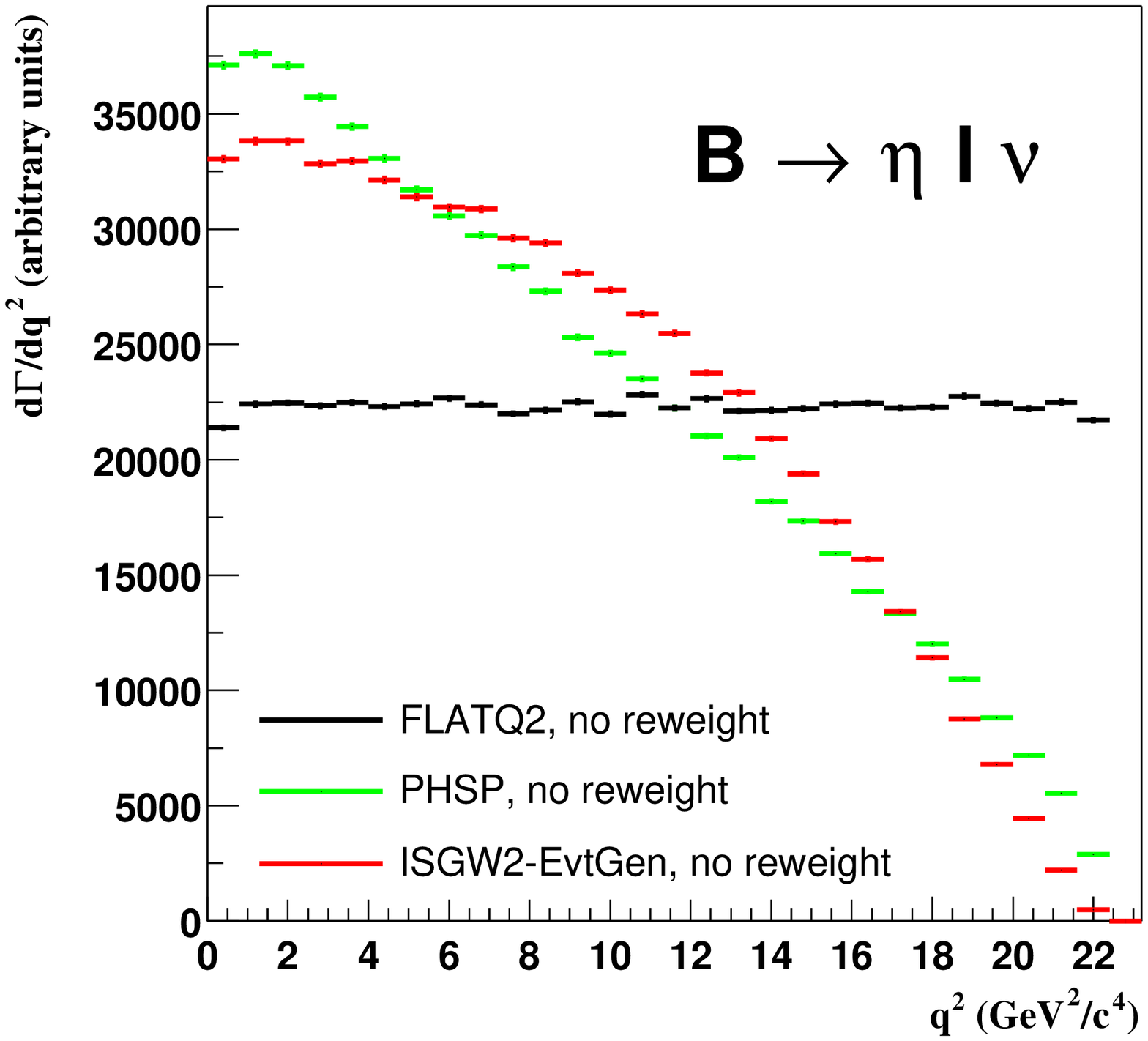}}
\resizebox{0.5\textwidth}{!}{\includegraphics{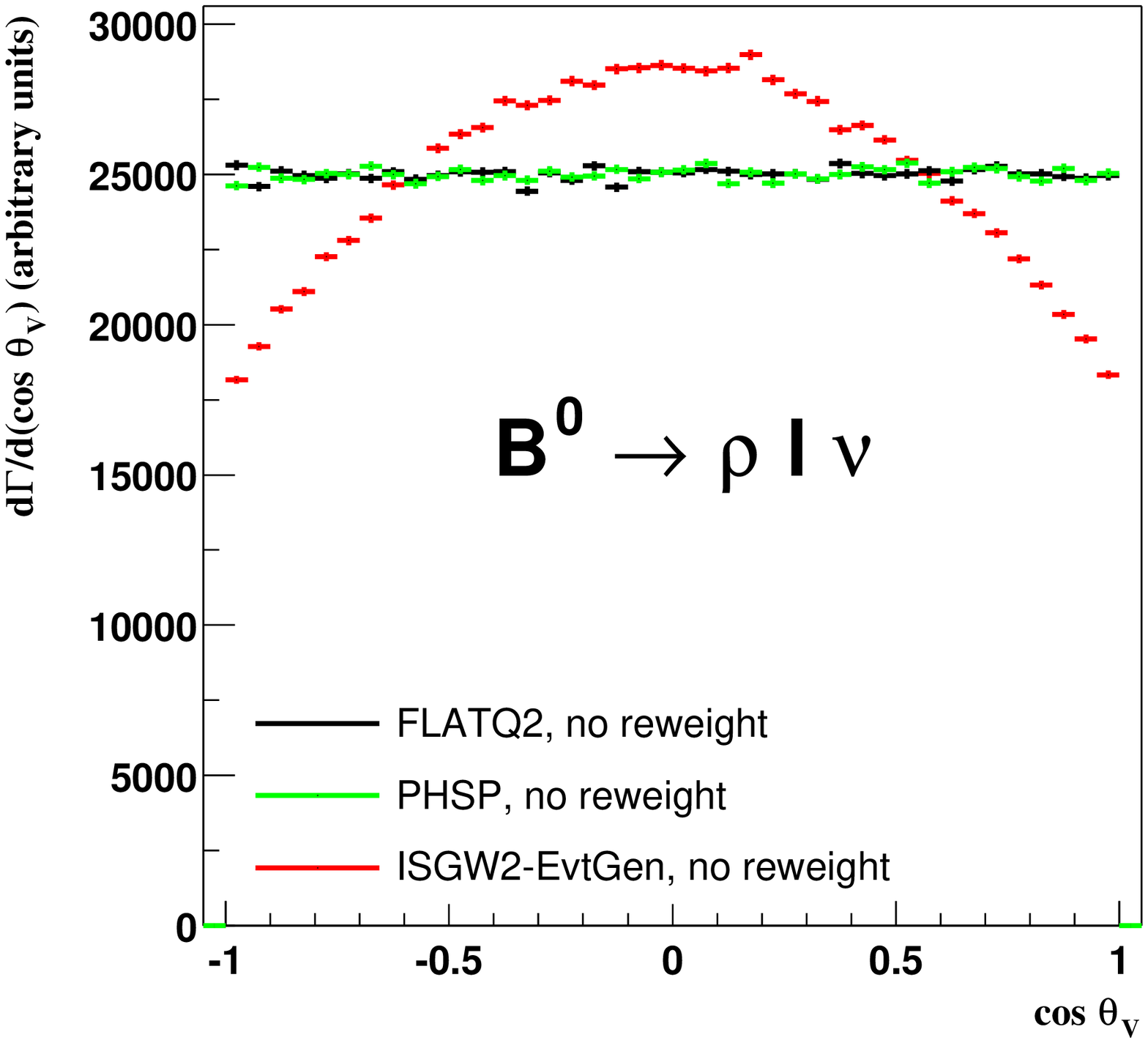}}}
\caption[Generator properties: $q^2$ (vectors) and $\cos(\theta_{V})$]
{ \label{fig4} Events generated with the FLATQ2 (black), PHSP (pale grey) and 
ISGW2 (dark grey) generators: $q^2$ distributions for unweighted $B 
\rightarrow \eta \ell \nu$ decays (left panel); $\cos\theta_{V}$ distributions
for unweighted $B^0 \rightarrow \rho \ell \nu$ decays (right panel)}
\end{center}
\end{figure*}

\section{Validation of the reweighting technique and its software tool}
\label{sec:validation}

In most simulations, the fully simulated events take into account the lepton's 
Final State Radiation (FSR). However, the effect of FSR is not included in the 
reweighting formulas presented in Sect. \ref{Rew}. As a consequence, the 
XSLEvtFFWeight tool could yield the wrong weights unless corrected for FSR. We 
have found that the easiest way around this problem is to use the LorentzVector 
of the lepton calculated from the vectors of the other particles of the decay as 
those are not affected by FSR. This solution turns out to be very effective even 
in cases where the FSR modifies the lepton's energy by several GeVs. As long as 
the same lepton is used to compute the kinematic angles and the event weight, the
effect of the FSR on the form factor reweighting is negligible.  

 The histograms shown in this section were calculated with such lepton 
LorentzVectors. They are extracted from 1 million entries generated with no 
detector simulation. This generator information is what is used with full Monte 
Carlo simulation in real physics analyses. Only a small sample is presented in 
this section. More can be found in Ref. \protect\cite{cote}.

Our reweighting software has not been tested on any $B \rightarrow X_c \ell\nu$
decay but there is no reason to believe that the results will be any different
from those obtained with the $B \rightarrow X_u \ell\nu$ decays.

\subsection{Properties of the generators FLATQ2, PHSP and ISGW2}
\label{GenProp}

The properties of the three generators used in this work, and calculated with the
XSLKin tool, are illustrated by the distributions displayed in Fig. \ref{fig4}. 
The distributions for all the decay modes of interest have been investigated 
\protect\cite{cote} and found to be as expected. Fig. \ref{fig4} (left panel) 
shows that, both, the PHSP and ISGW2 generators yield low statistics at high 
$q^2$. This makes precise efficiency corrections difficult in this important 
region (particularly important for lattice QCD tests). Fig. \ref{fig4} (right 
panel) shows that the PHSP and FLATQ2 $\cos \theta_V$ distributions are 
identical. This is also true for the $\cos\theta_{\ell}$ and $\chi$ angle 
distributions \cite{cote}. Note that since the FLATQ2 generator yields flat 
distributions for all the variables ($\theta_{\ell}, \theta_V, \chi, q^2$) of 
interest, it allows a smooth reweighting over the full phase space. It is thus 
ideally suited to evaluate and correct the efficiencies as a function of these 
variables. To determine the analysis cuts required in realistic physics 
simulation, the FLATQ2 and PHSP generated events have to be reweighted. 

\begin{figure*}[ht]
\begin{center}
\mbox{
\resizebox{0.35\textwidth}{!}{\includegraphics{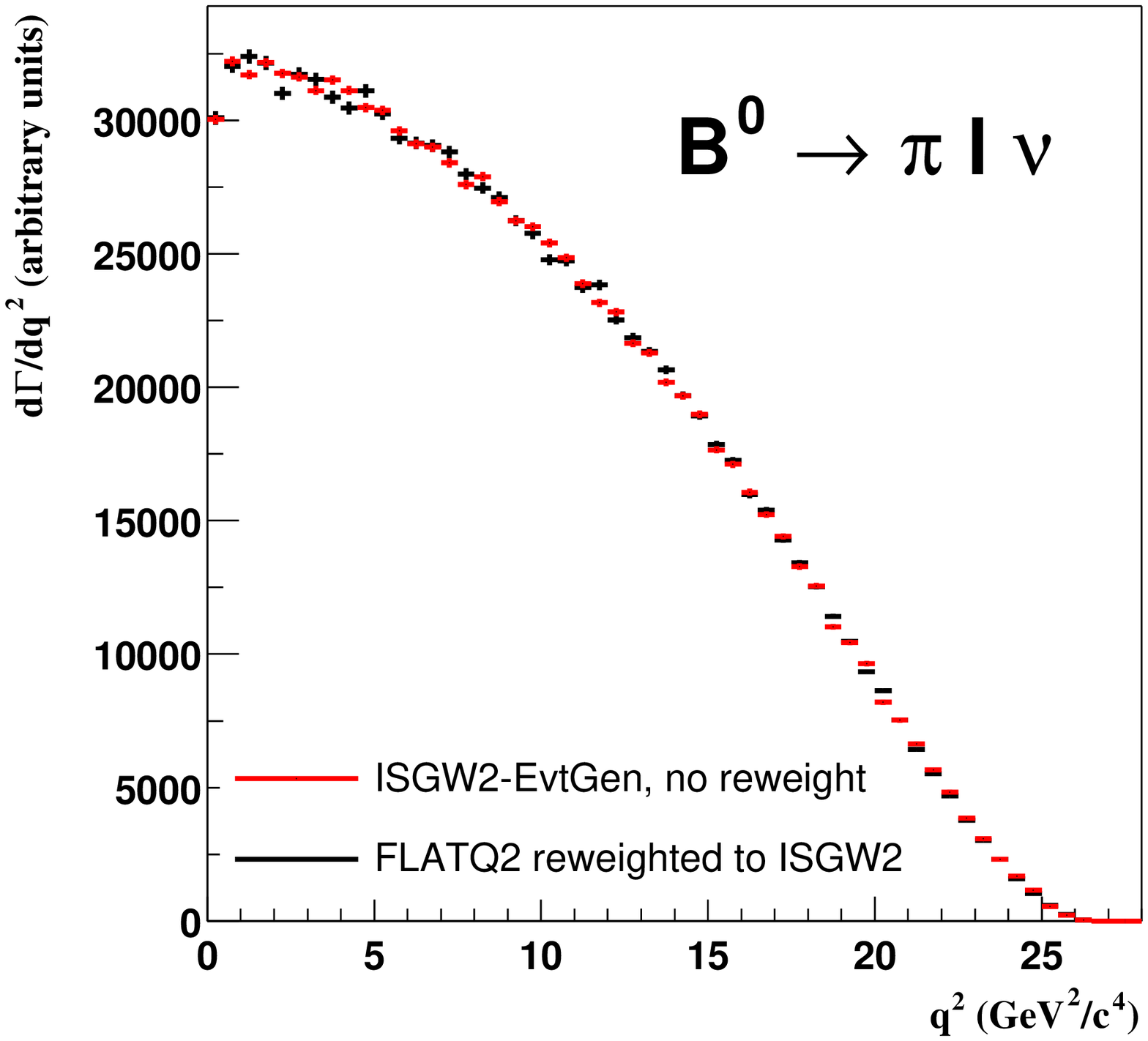}}
\resizebox{0.35\textwidth}{!}{\includegraphics{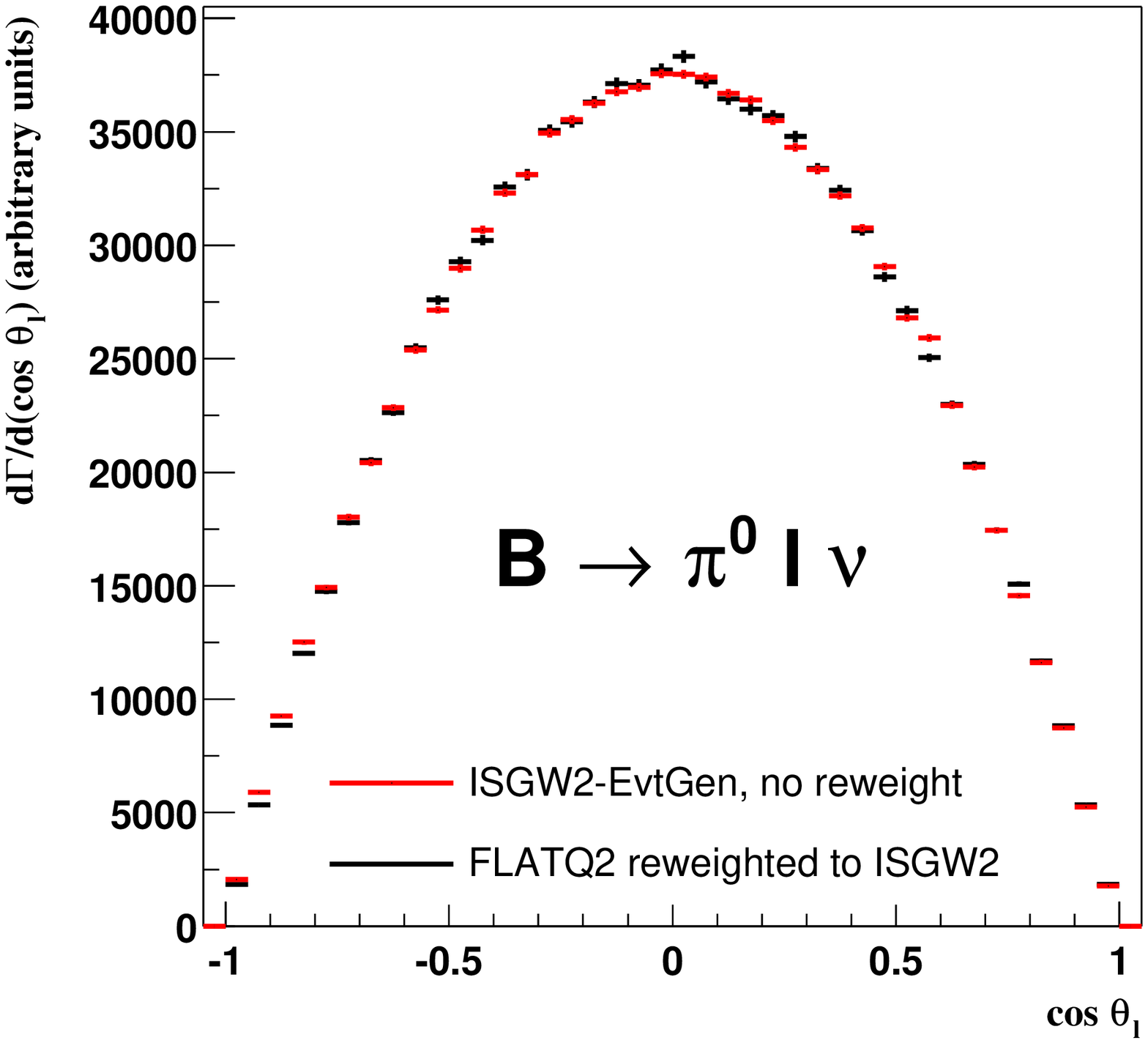}}
\resizebox{0.35\textwidth}{!}{\includegraphics{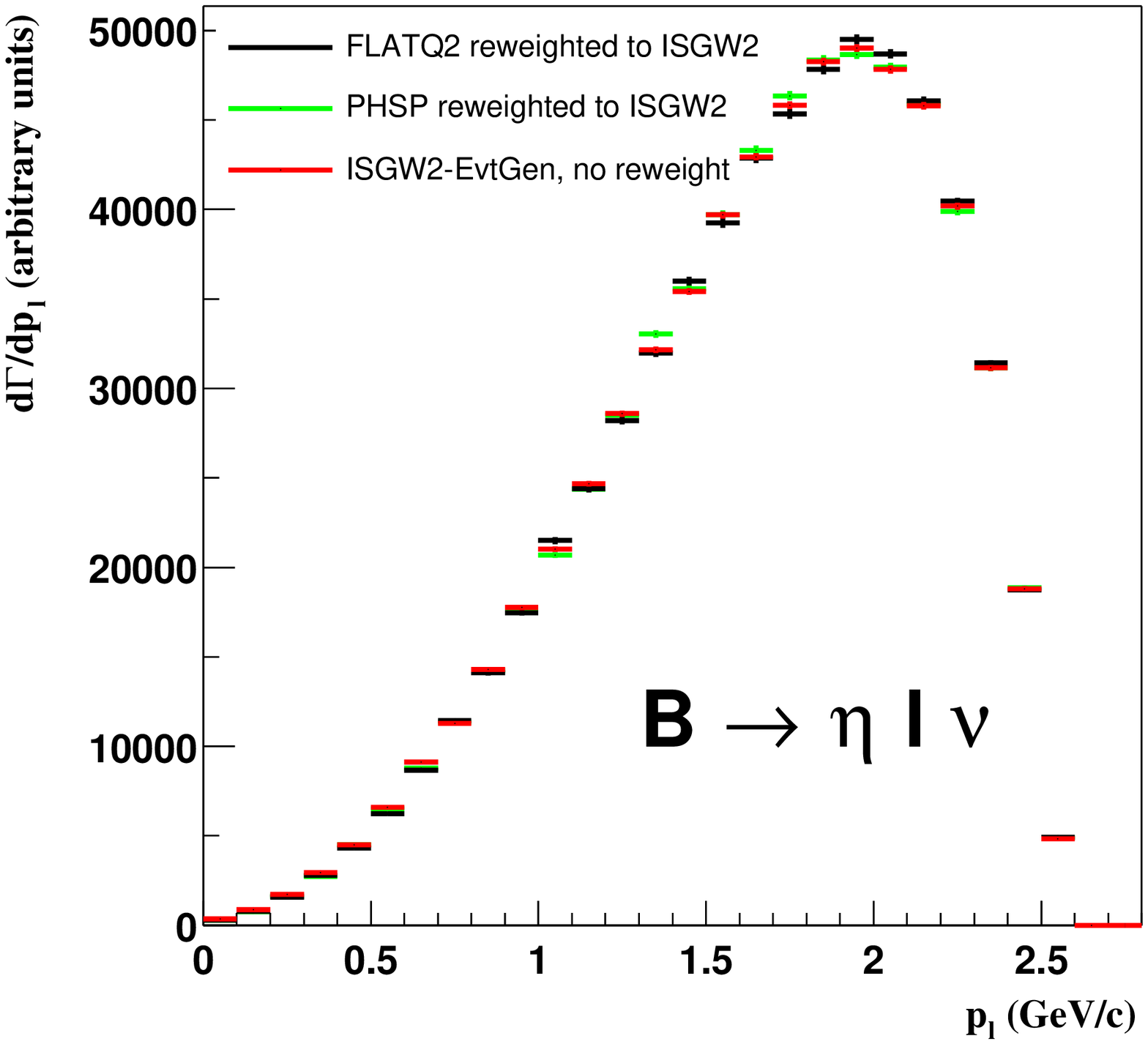}}}
\caption[Reweighting of B to pseudo-scalar decays]{\label{fig5}
Pseudo-scalar decays. Events generated with a FLATQ2 generator and reweighted to 
the ISGW2 form factor hypothesis (black), events generated with a PHSP generator 
and reweighted to the ISGW2 form factor hypothesis (pale grey) and unweighted 
events generated directly with a ISGW2 generator (dark grey): comparison of $q^2$
distributions for $B \rightarrow \pi^{+} \ell \nu$ decay (left panel); comparison
of $\cos\theta_{\ell}$ distributions for $B \rightarrow \pi^0 \ell \nu$ decays 
(center panel); comparison of $p_{\ell}$ distributions, in the true B frame, for 
$B \rightarrow \eta \ell \nu$ decays (right panel).}
\end{center}
\end{figure*}

\begin{figure*}[h]
\begin{center}
\mbox{
\resizebox{0.35\textwidth}{!}{\includegraphics{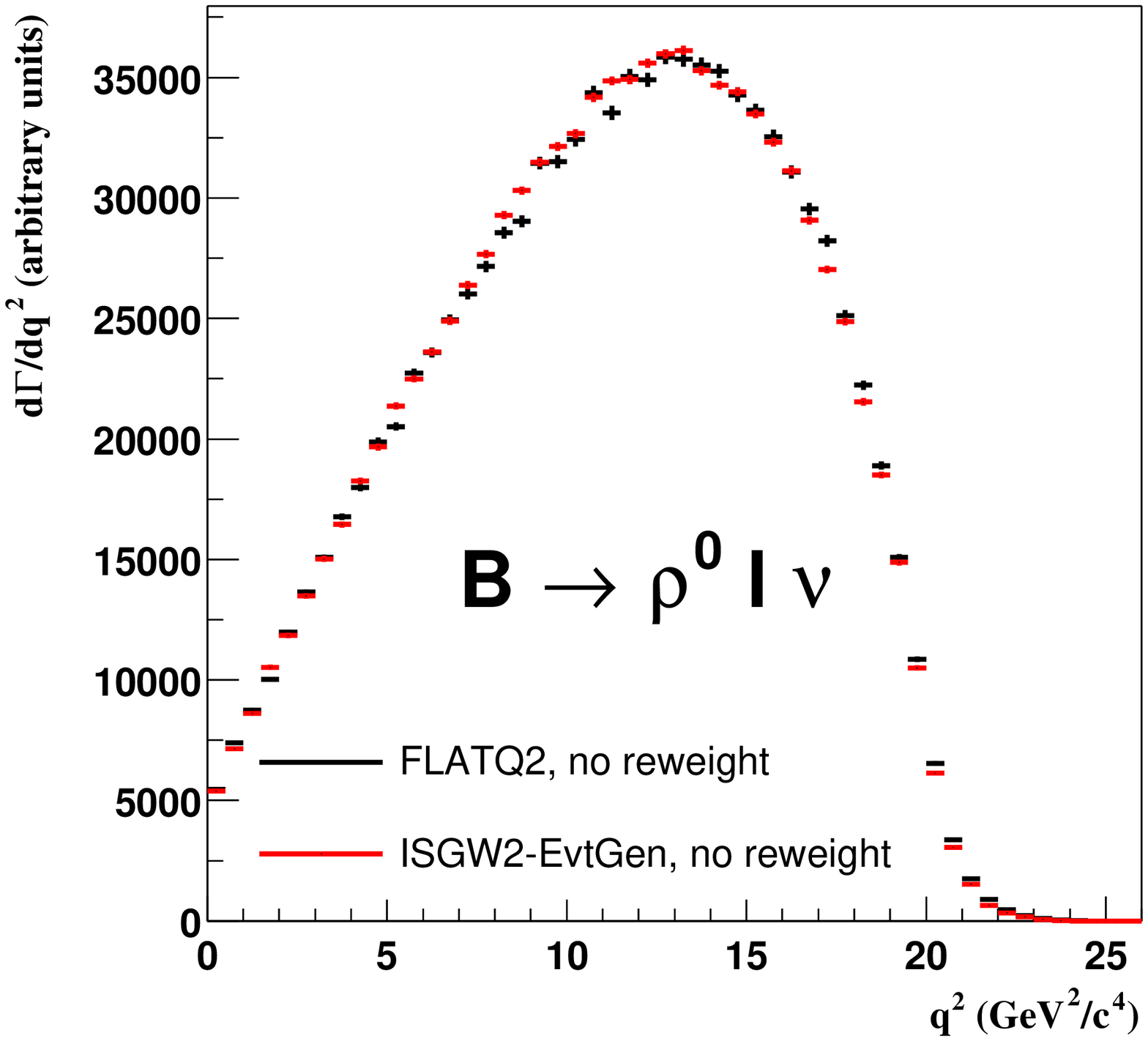}}
\resizebox{0.35\textwidth}{!}{\includegraphics{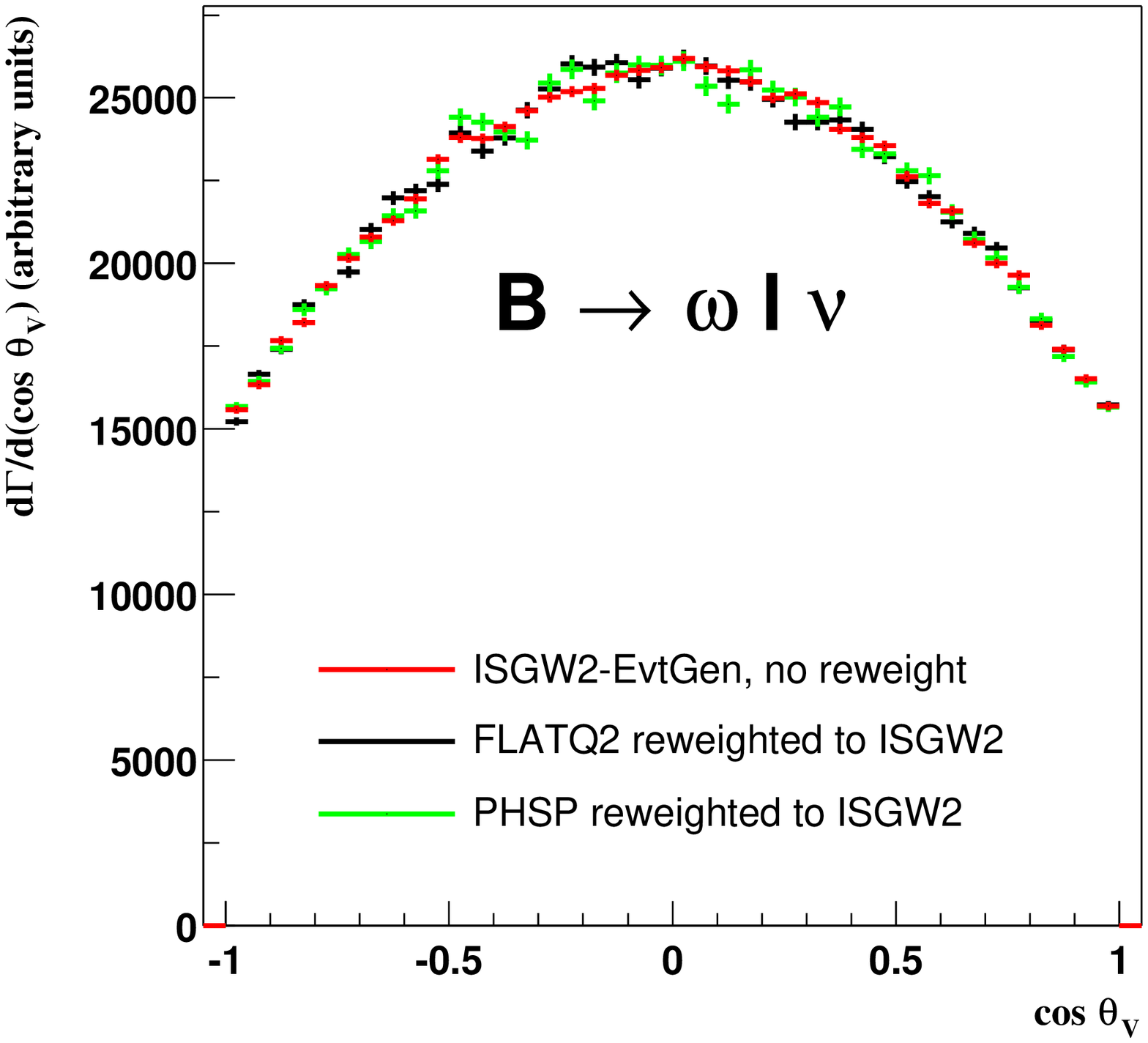}}
\resizebox{0.35\textwidth}{!}{\includegraphics{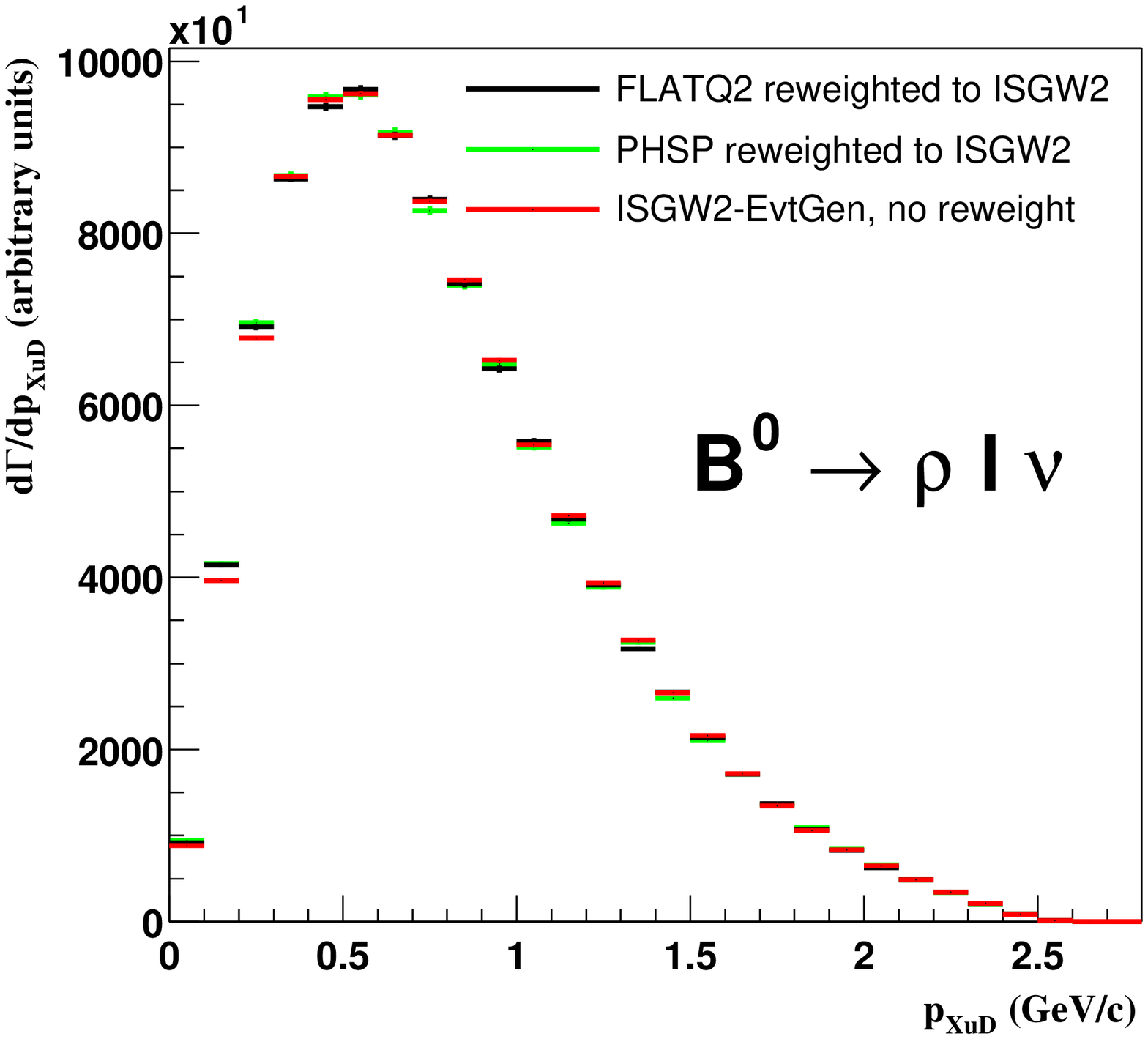}}}
\caption[reweighting of B to vector meson decays]{ \label{fig6} 
Vector meson decays. Events generated with a FLATQ2 generator and reweighted to 
the ISGW2 form factor hypothesis (black), events generated with a PHSP generator 
and reweighted to the ISGW2 form factor hypothesis (pale grey) and unweighted 
events generated directly with a ISGW2 generator (dark grey): comparison of $q^2$
distributions for $B \rightarrow \rho^0 \ell \nu$ decays (left panel); comparison
of $\cos\theta_V$ distributions for $B \rightarrow \omega \ell \nu$ decays 
(center panel); comparison of $p_{XuD}$ distributions, in the true B frame, for 
$B \rightarrow \rho^+ \ell \nu$ decays (right panel). $p_{XuD}$ is the momentum 
of any daughter of the $X_u$ meson. }
\end{center}
\end{figure*}

\subsection{Reweighting of B to pseudo-scalar meson decays}
\label{ValidationScalar}

Note that because of the three-level hierarchy of the XSLEvtFFWeight tool, it 
is sufficient to fully validate the software with only one generator. For this
purpose, the ISGW2 generator was used. In this way, the XSLKin tool (Sect. 
\ref{XSLKin}) and the classes XSLEvtFFWeight, XSLPseudoScalarFF, 
XSLVectorFF, XSLScalarISGW2 and XSLVectorISGW2 (Sect. \ref{XSLEvtFFWeight})
are all validated. In the implementation of any new form factor model, it is then
sufficient to ensure the correctness of the new form factor equations at the 
third level of the XSLEvtFFWeight tool (see Sect. \ref{newff}).

As can be seen in Fig. \ref{fig5}, there is an excellent match between the 
distributions generated directly with the ISGW2 generator and those generated 
with the FLATQ2 or PHSP generators, and then reweighted to the ISGW2 form factor 
hypothesis with our software.

\subsection{Reweighting of B to vector meson decays}
\label{ValidationVec}

Like for the reweighted pseudo-scalar meson decay results, there is the same 
excellent match (Fig. \ref{fig6}) when reweighting is applied to the 
distributions generated with the FLATQ2 or PHSP generators for vector meson 
decays.

\clearpage
\subsection{Validity of the massless lepton approximation}
\label{massless}

\begin{figure}[h]
\begin{center}
\mbox{
\resizebox{0.45\textwidth}{!}{\includegraphics{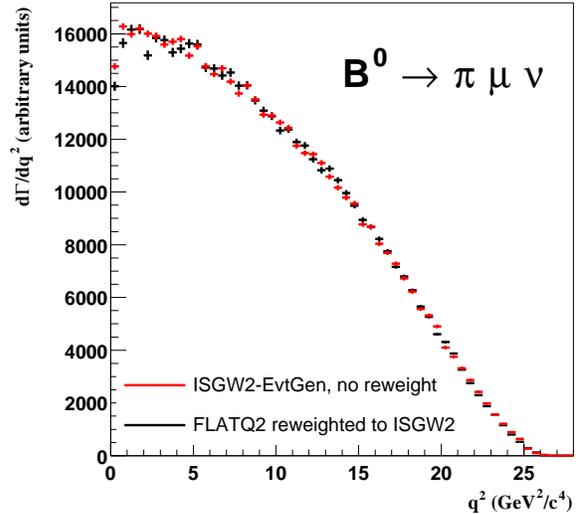}}}
\caption[$B \rightarrow S \ell \nu$ validation: $q^2$ for muons only]
{ \label{fig7}
Comparison of $q^2$ distributions for $B \rightarrow \pi \mu \nu$ decays only. 
Events generated with a FLATQ2 generator and reweighted to the ISGW2 form 
factor hypothesis are shown in black while the unweighted events generated 
directly  with a ISGW2 generator are shown in dark grey.}
\end{center}
\end{figure}

Comparing various distributions generated with a standard ISGW2 generator to 
those generated with our form factor reweighting software allows us to test the 
validity of the massless approximation used in our reweighting software. This is 
the case since the standard code uses the exact formulas to compute the 
differential decay rates. Since the comparisons, done separately for electrons 
and muons, show the same good match, it can be concluded that the massless 
approximation is indeed justified for both, electrons and muons. An exemple of 
the quality of the match is displayed in Fig. \ref{fig7}.

\section{Improvements in measurements of $|V_{ub}|$}
\label{multimodels}

\begin{figure*}[ht]
\begin{center}
\mbox{
\resizebox{0.35\textwidth}{!}{\includegraphics{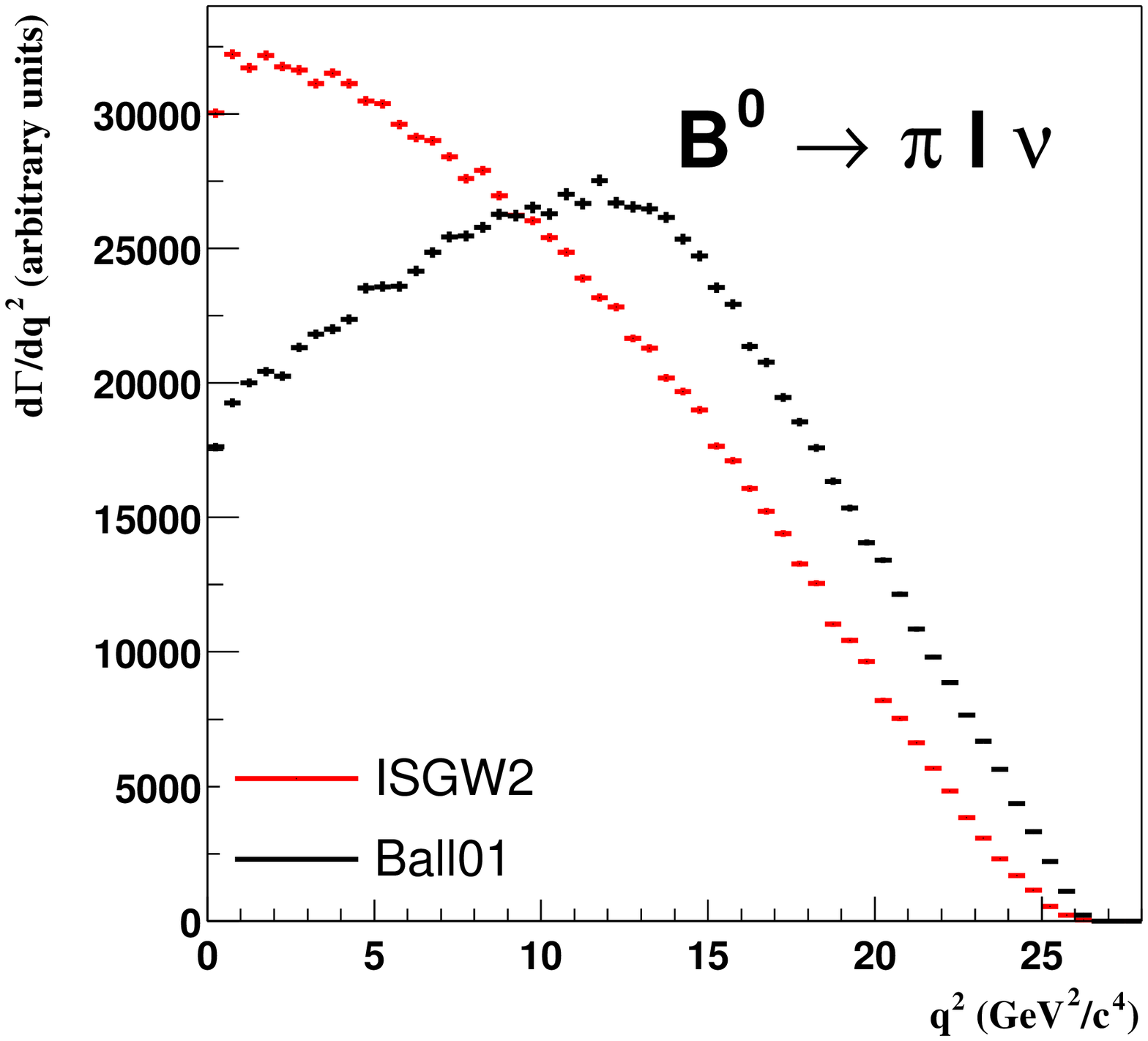}}
\resizebox{0.35\textwidth}{!}{\includegraphics{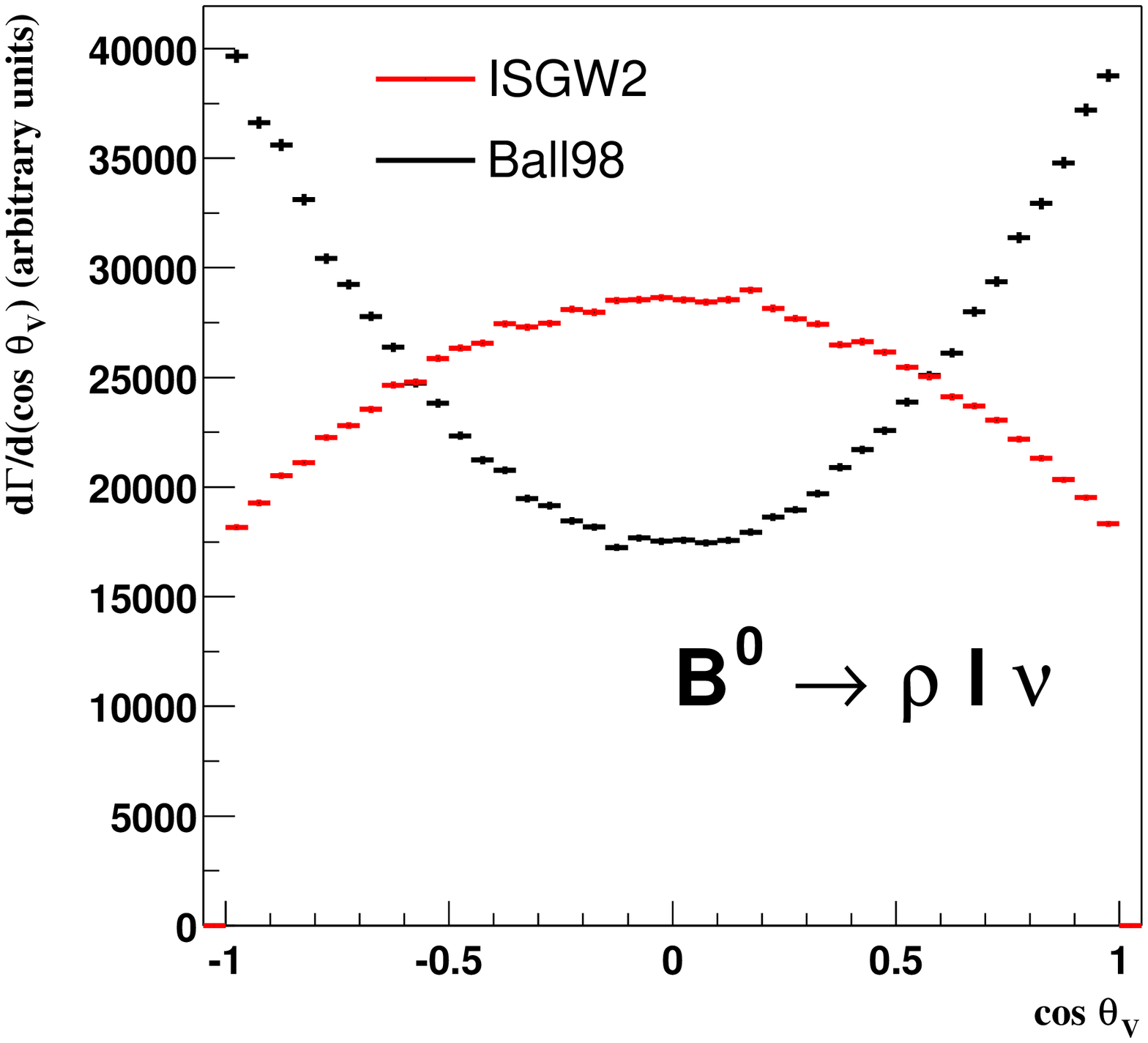}}
\resizebox{0.35\textwidth}{!}{\includegraphics{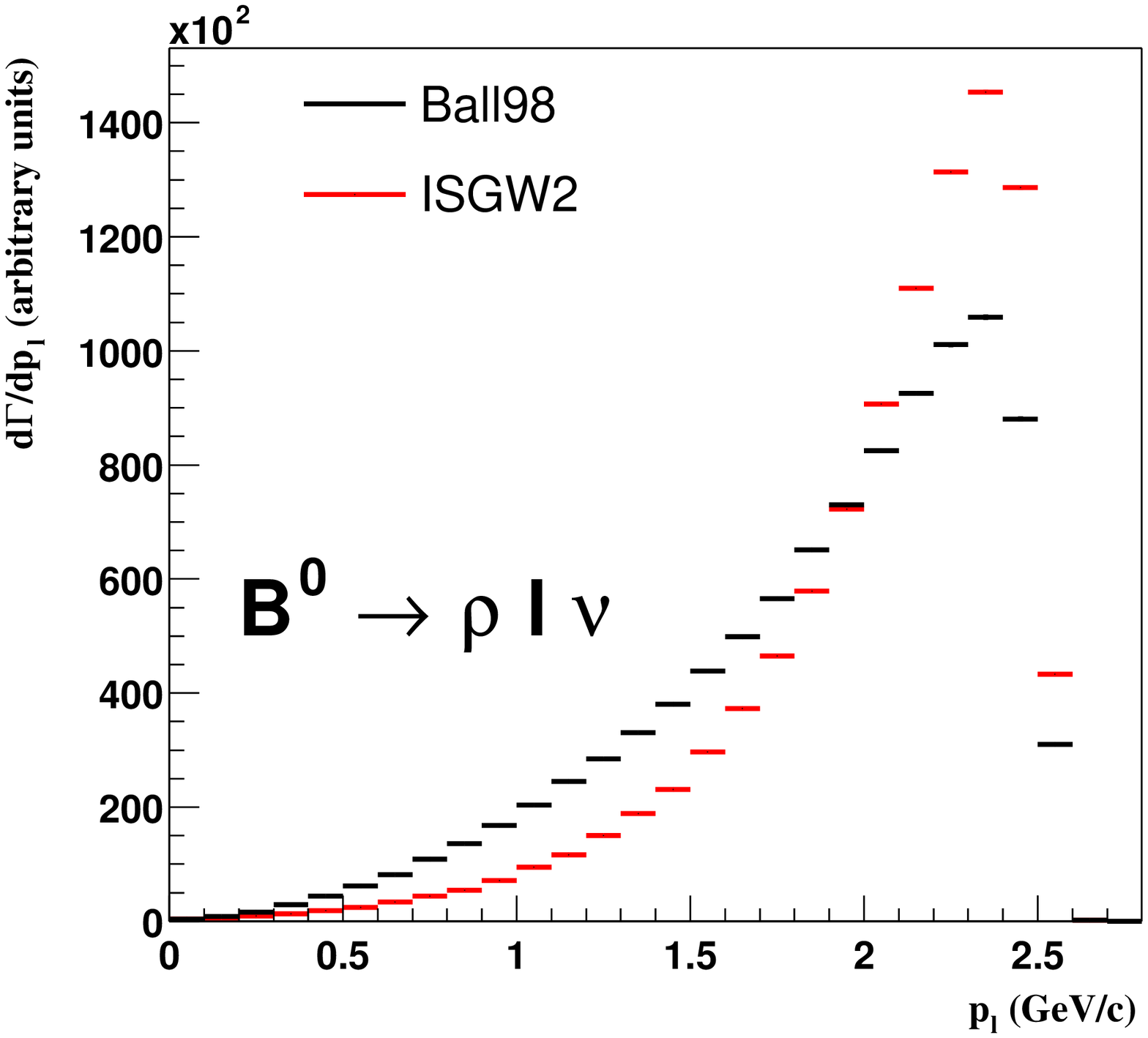}}}
\caption[ISGW2 vs Ball01: $q^2$, $\cos\theta_V$, $P_{\ell}]{ \label{fig8} 
Events reweighted with the Ball01-LCSR form factor hypothesis (black), unweighted
events generated directly with the ISGW2 form factor hypothesis (dark grey): 
comparison of $q^2$ distributions for $B \rightarrow \pi^{+} \ell \nu$ decays 
(left panel); comparison of $\cos\theta_V$ distributions for $B \rightarrow 
\rho^{+} \ell \nu$ decays (center panel); comparison of $p_{\ell}$ distributions 
for $B \rightarrow \rho^{+} \ell \nu$ decays (right panel).}
\end{center}
\end{figure*}

\begin{figure*}[ht]
\begin{center}
\mbox{
\resizebox{0.5\textwidth}{!}{\includegraphics{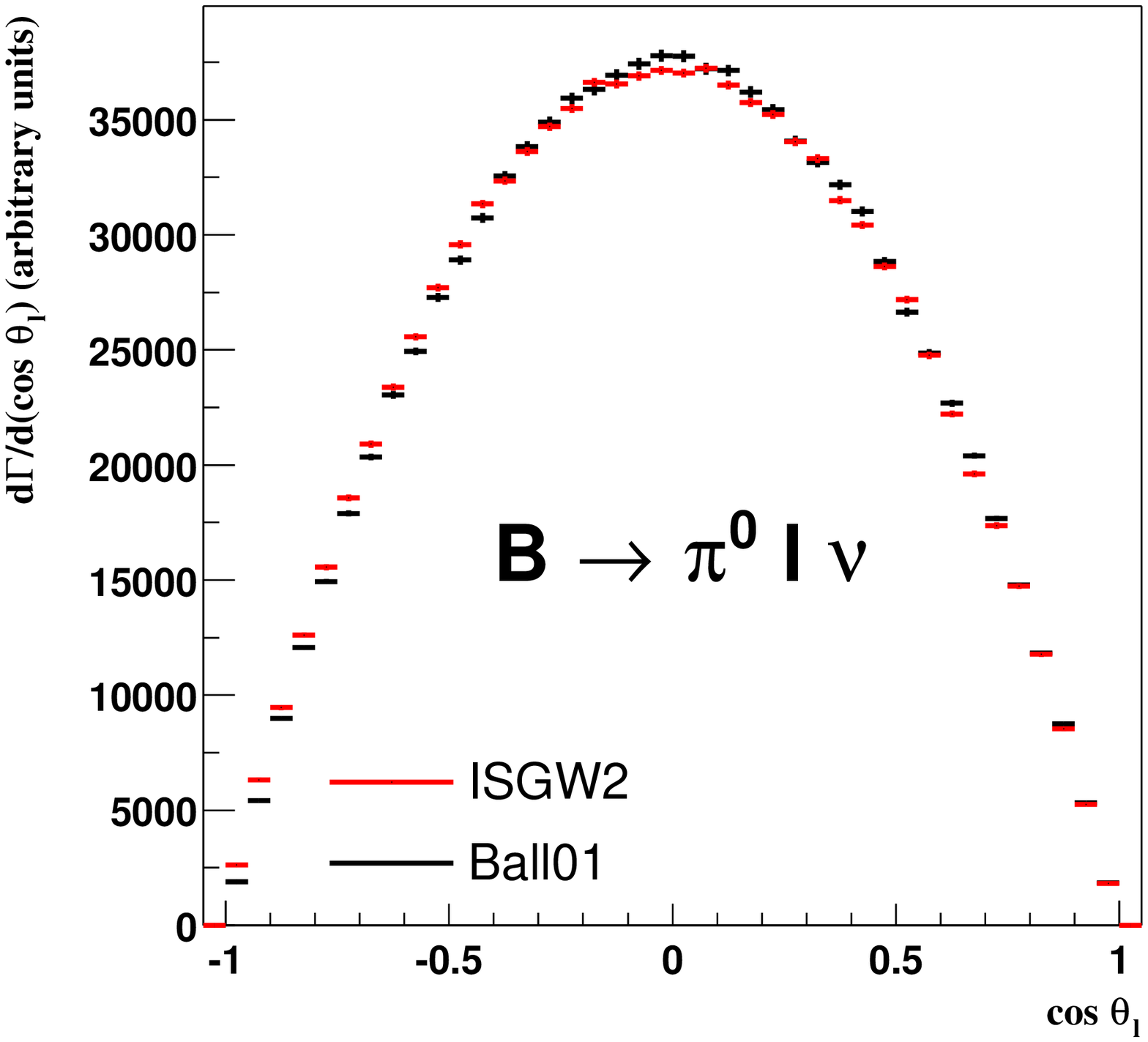}}
\resizebox{0.5\textwidth}{!}{\includegraphics{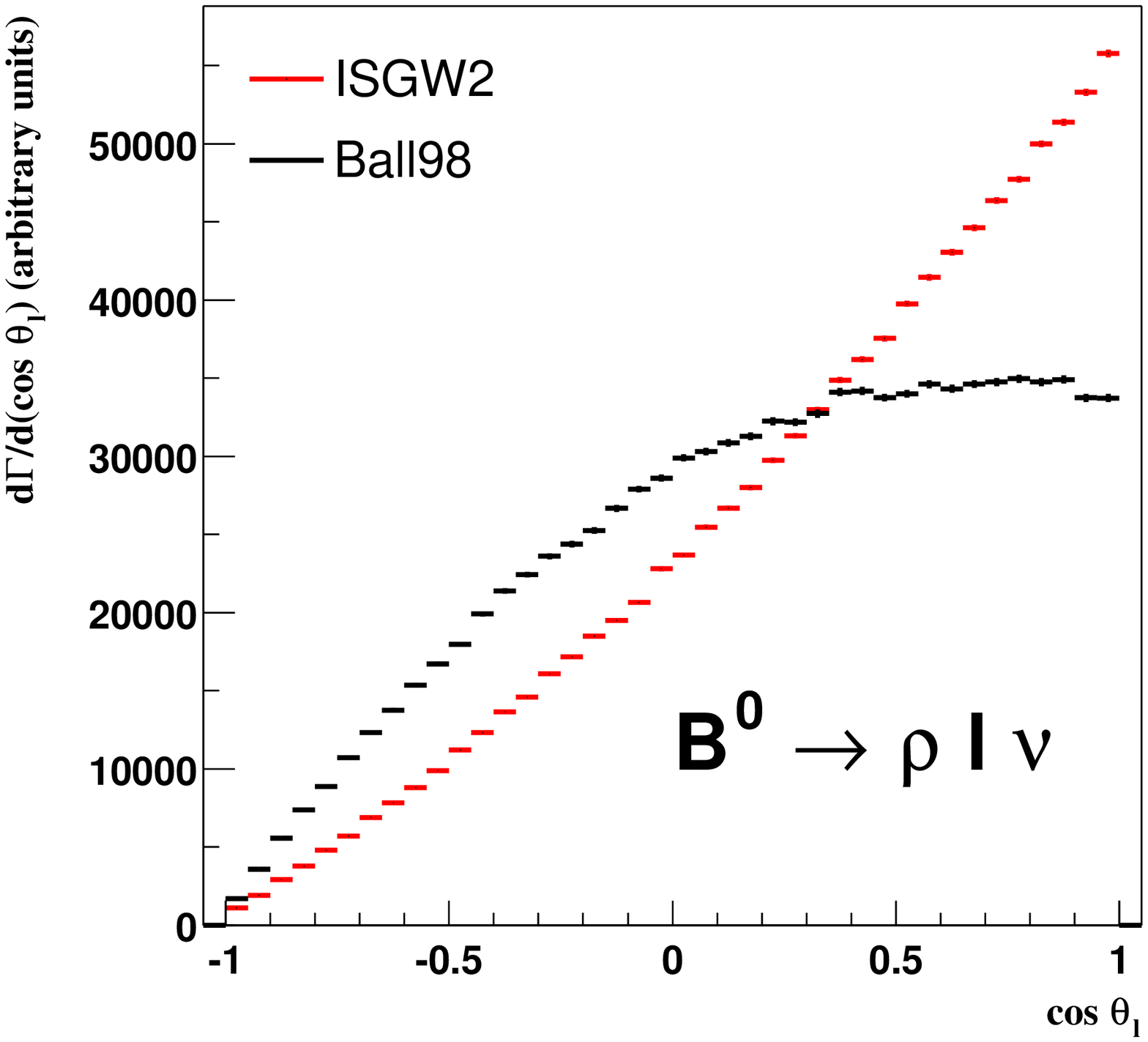}}}
\caption[ISGW2 vs Ball01: $\cos(\theta_{\ell})$]{ \label{fig9} 
Events reweighted with the Ball01-LCSR form factor hypothesis (black), unweighted
events generated directly with the ISGW2 form factor hypothesis (dark grey): 
comparison of $\cos\theta_{\ell}$ distributions for $B \rightarrow \pi^{0} \ell 
\nu$ pseudo-scalar meson decays (left panel); comparison of $\cos\theta_{\ell}$ 
distributions for $B \rightarrow \rho^{+} \ell \nu$ vector meson decays (right 
panel).}
\end{center}
\end{figure*}

With our reweighting software, it is now easy to investigate the predictions of 
various form factor models, and to evaluate their impact on the experimental 
study of $B \rightarrow X_u \ell \nu$ decays. So far, two form factor models
for pseudo-scalar and vector meson decays have been fully implemented in our
software: the ISGW2 \protect\cite{isgw2} model (used in the validation of the 
reweighting technique and its software tool) and a LCSR model from P. Ball et al.
\protect\cite{ball98}\protect\cite{ball01}. Typical distributions for $q^2$ 
\footnote{$q^2$ is uniquely related to $p_{Xu}$ in the $B$ frame as shown by Eq. 
\ref{pXq2}.}, $\cos \theta_V$ and $p_{\ell}$ (Fig. \ref{fig8}) deduced from these
two models display a large difference while the $\cos\theta_{\ell}$ distributions
for pseudo-scalar meson decays (Fig. \ref{fig9}, left panel) are not 
model-dependent, but those for vector meson decays (Fig. \ref{fig9}, right panel)
are. This significant model-dependence shows why the values of the branching 
fraction and of the CKM matrix element $|V_{ub}|$, extracted from the study of 
exclusive semileptonic B meson decays, have such large theoretical errors. 

The most precise value to date of $|V_{ub}|$ obtained from the analysis of 
exclusive semileptonic B decays is the one given in Ref. \protect\cite{pdg}:

$$
 |V_{ub}| = (3.27 \pm 0.13 \pm 0.19^{+0.51}_{-0.45})\times 10^{-3}
$$
where the first error is statistical, the second systematic and the third 
theoretical. It is clear that the theoretical error, of the order of 15\%, due
to form factor uncertainties, is by far the largest error. With our reweighting
software tool, using our FLATQ2 generator, specially useful to determine the
critical efficiencies of our experimental cuts as a function of $q^2$, we aim 
to reduce the theoretical error to approximately 5\%. The integrated luminosity
required to achieve this goal depends on the technique used to reconstruct the
exclusive decay channel. With the neutrino reconstruction technique 
\protect\cite{babar1}, and the more than 250 $fb^{-1}$, collected at the 
$\Upsilon(4S)$ resonance in both BaBar and Belle, and of the order of 
500 $fb^{-1}$ expected within the next two years, a smaller theoretical error 
should be possible in the near future. With the recoil techniques 
\protect\cite{babar2}, and for the other exclusive $\rho\ell\nu$, 
$\omega\ell\nu$, $\eta\ell\nu$ and $\eta^{\prime}\ell\nu$ decay channels, a few 
additional hundreds of $fb^{-1}$ will be required to extract a similar 
theoretical precision on $|V_{ub}|$.

With the soon to be available data from both BaBar and Belle, and with the
neutrino reconstruction technique, a statistical error below 1\% is possible. To
achieve a theoretical error at a few percent level will require tight constraints
on the form factors as a function of $q^2$. This can be accomplished by two 
methods. 

 In the first method, the measured differential decay distributions are compared 
to the corresponding ones predicted by various form factor models, as shown e.g. 
in Fig. 8. Our reweighting software tool will play a crucial role in these 
comparisons since it will allow a quick and easy test of all models without
having to regenerate a complete MC production for each model. In this method, the
efficiencies of the analysis cuts are computed with the model used to generate 
the distributions. With very high statistics, the method should allow us to keep 
a single form factor model as being the one closest to reality, thereby reducing 
the theoretical error on $|V_{ub}|$. 

In the second method, the form factors will be measured directly with minimal
model dependence. For example, for pseudo-scalar decays, the measured 
distributions will be fitted to Eq. 10 where the form factor is given e.g. 
\protect\cite {ball04} by:
$$
 f^+(q^2) = \frac{r_1}{1-q^2/m^2} +
 \frac{r_2}{\left(1-q^2/m^2\right)^2} 
$$
where $r_1$, $r_2$ and $m$ are the parameters to be fitted. Other functions will 
have to be investigated to obtain the systematic error engendered by the use of 
such functions. The small model dependence comes from the differential 
efficiencies, $\epsilon(q^2(,\theta_{\ell},\theta_V,\chi))$, of the analysis cuts
used to generate the measured distributions. To evaluate these efficiencies, the 
four kinematical variable distributions have to be binned, and the efficiencies 
are then determined for each bin. 

 In the limit of infinitesimal bins, the efficiency $\epsilon$ is independent of
any model, as shown in Ref. [1]. But, of course, in any practical analysis, with
finite statistics and limited kinematical variable resolutions, the bin size will
be finite. This introduces a small model dependency which must be taken into 
account. It can be limited by the appropriate choice of analysis cuts since it
can be shown \protect\cite{cote} that if the cuts are not correlated with the 
kinematic variables, then the efficiency is model-independent. The choice is 
greatly facilitated by the use of the FLATQ2 generator to determine the values of
$\epsilon$, especially at very high values of $q^2$. It is also facilitated by 
the use of the form factor reweighting tool to select the experimental cuts to 
limit the correlation and thus the model dependence. The generator, combined with
the software tool, will also yield a precise value of the small residual
uncertainty due to the finite size of the bins. Once the parameters of the form 
factors are determined, they can be directly compared to any model predictions,
thereby reducing the number of models and thus the theoretical uncertainty on
$|V_{ub}|$.   

\section{Conclusions}
\label{con}

A form factor reweighting technique and its software tool have been presented in
this paper. Both have been validated with the ISGW2 form factor model as 
illustrated by the excellent match between the distributions generated with the 
FLATQ2 or PHSP generator, reweighted to the ISGW2 form factor hypothesis, and 
those generated directly with a ISGW2 generator. The object-oriented design of 
the software tool allows an easy and reliable implementation of any new form 
factor model, while optimizing the required CPU resources. The large differences,
easily observed with our tool, in the distributions predicted by the ISGW2 and 
LCSR models for exclusive $B \rightarrow X_u \ell \nu$ decays show that a study 
of these decays will be very valuable in extracting the values of the form 
factors. Our work leads us to expect that both, our novel FLATQ2 generator and 
form factor reweighting tool, will play a key role in the next generation of 
exclusive $|V_{ub}|$ measurements, with largely reduced errors.

\bigskip

{\it Acknowledgements} We are grateful to M. S. Gill for providing Fig. 1. We 
wish to thank SLAC for its kind hospitality. This work was financially supported 
by NSERC (Canada).



\end{document}